\documentclass[prd,twocolumn,showpacs,preprintnumbers,amsmath,amssymb,superscriptaddress,floatfix,nofootinbib]{revtex4-2}

\usepackage{graphicx}
\usepackage{bm}
\usepackage{amsmath}
\usepackage{amsfonts}
\usepackage{amssymb}
\usepackage{color}
\usepackage{multirow}
\usepackage{subfigure}
\usepackage{txfonts}
\usepackage[colorlinks, citecolor=blue,anchorcolor=red,menucolor=red, linkcolor=red,filecolor=red,runcolor=red,urlcolor=blue,frenchlinks=red]{hyperref}

\begin{document}
	\title{Theoretical study of the $\psi(3686)\to\bar{p}K^*(892)\Sigma^+$ and the $\psi(3686)\to\bar{p}K^+\Sigma^*(1385)$ reactions}

    \author{Wen-Tao Lyu}\email{lvwentao9712@163.com}
    \affiliation{School of Physics, Zhengzhou University, Zhengzhou 450001, China}
    \affiliation{Departamento de Física Teórica and IFIC, Centro Mixto Universidad de Valencia-CSIC Institutos de Investigación de Paterna, 46071 Valencia, Spain}
    \vspace{0.5cm}
    
    \author{Fernando C\'esar Sobrinho}
    \affiliation{Instituto de F\'\i{}sica, Universidade de S\~ao Paulo,
    	Rua do Mat\~ao 1371, CEP 05508-090, S\~ao Paulo, SP, Brazil}
    \affiliation{Departamento de Física Teórica and IFIC, Centro Mixto Universidad de Valencia-CSIC Institutos de Investigación de Paterna, 46071 Valencia, Spain}
    
    \author{Man-Yu Duan}\email{duanmy@ppsuc.edu.cn}
    \affiliation{School of Information Network Security, People's Public Security University of China, Beijing 100038, China}\vspace{0.5cm}
    
    \author{Eulogio Oset}\email{oset@ific.uv.es}
    \affiliation{Departamento de Física Teórica and IFIC, Centro Mixto Universidad de Valencia-CSIC Institutos de Investigación de Paterna, 46071 Valencia, Spain}
    \affiliation{Department of Physics, Guangxi Normal University, Guilin 541004, China
    }

\begin{abstract}
	
Motivated by the recent BESIII measurement, we investigate the $\psi(3686)\to\bar{p}K^*(892)\Sigma^+$ and the $\psi(3686)\to\bar{p}K^+\Sigma^*(1385)$ decays in a coupled-channel framework, incorporating $S$-wave final-state interactions in the vector-meson--antibaryon and pseudoscalar-meson--antibaryon subsystems, respectively. The coupled-channel amplitudes dynamically generate the $\Sigma(1830)$ resonance in the vector-baryon sector and the $\Sigma(1430)$ in the pseudoscalar-baryon sector. A fit with three free parameters provides a reasonable overall description for $\psi(3686)\to\bar{p}K^*(892)\Sigma^+$. In particular, the $\Sigma(1830)$ gives rise to a pronounced near-threshold enhancement in the $\bar{p}K^*(892)$ mass distribution. For the $\psi(3686)\to\bar{p}K^+\Sigma^*(1385)$ decay, a fit with three free parameters also provides a reasonable description, where the $\Sigma(1430)$ gives rise to a pronounced near-threshold enhancement in the $\bar{p}K^+$ distribution. Furthermore, we estimate the theoretical uncertainties of the invariant mass distributions using a parametric bootstrap procedure. Higher-precision measurements of these decays would help constrain the final-state interactions and clarify the properties of the $\Sigma(1430)$ and $\Sigma(1830)$ resonances.

\end{abstract}
	
	%\keywords{ }
	\date{\today}
	
	\maketitle
	
\section{Introduction}\label{sec1}

The light ground-state baryon octet and decuplet are well established, but the spectrum of low-lying excited baryons with spin-parity $J^P=1/2^-$ remains less well understood. In the $\Sigma$ sector, the 2026 edition of the Particle Data Group (PDG) review assigns a one-star rating to the $\Sigma(1620)$ with $J^P=1/2^-$~\cite{ParticleDataGroup:2026mpi}, indicating that its existence is not yet firmly established~\cite{Wang:2024jyk,Sarantsev:2019xxm}. Following earlier theoretical studies~\cite{Xie:2017xwx,Lyu:2024qgc}, BESIII reported evidence for a $J^P=1/2^-$ $\Sigma$ state with a mass near $1380\,\mathrm{MeV}$ in the decay $\Lambda_c^+\to\Lambda\pi^+\eta$, with a statistical significance exceeding $3\sigma$~\cite{BESIII:2024mbf}. Other theoretical studies have also examined a possible $\Sigma(1380)$ contribution~\cite{Wu:2009nw,Li:2024rqb,Lyu:2026ack,Lyu:2026ums}.

Another one is $\Sigma(1430)$ state. The $\Sigma(1430)$ was predicted theoretically in the study of the meson baryon interaction in the context of the chiral unitary approach~\cite{Oller:2000fj,Jido:2003cb}, and it appeared as a cusp in the $\bar{K}N$ threshold, or as a slightly bound state depending on some small differences in the input of the interaction. The study of $\pi\Sigma$ photoproduction reactions of Ref.~\cite{CLAS:2013rjt} in Ref.~\cite{Roca:2013cca} also showed the appearance of the $\Sigma(1430)$ state as a cusp at the $\bar{K}N$ threshold. Other works along the lines of the chiral unitary approach in coupled channels come to support these findings~\cite{Oset:2001cn,Oller:2006jw,Garcia-Recio:2002yxy,Lutz:2001yb,Guo:2012vv,Kamiya:2016jqc,Khemchandani:2018amu}. In Ref.~\cite{Xie:2018gbi}, the $\Lambda_c^+ \to \pi^+\pi^0\pi^-\Sigma^+$ reaction was analyzed in terms of a triangle singularity that reinforced the production of the $\Sigma(1430)$ state. Also, in Ref.~\cite{Xie:2018gbi}, it was suggested to look at the $\Lambda_c^+ \to \pi^+\pi^+\pi^-\Lambda$ reaction, where the signal for this resonance should be equally seen. The Belle Collaboration measured $\Lambda_c^+ \to \pi^+\pi^+\pi^-\Lambda$ recently~\cite{Belle:2022ywa} and found a clear signal at the $\bar{K}N$ threshold, both in the $\Lambda\pi^+$ and $\Lambda\pi^-$ mass distributions, providing the first clear evidence for the existence of this state. It is also shown that one cannot discriminate between having a resonance or a cusp at the $\bar{K}N$ threshold~\cite{Belle:2022ywa}. Yet, from our point of view, there is only a small and smooth transition from one scenario to the other by gradually changing the strength of the interaction. The $\Lambda_c^+ \to \pi^+\pi^+\pi^-\Lambda$ reaction was studied later in Ref.~\cite{Li:2025yad} with the same framework of Ref.~\cite{Xie:2018gbi}, and it was found that the signal and production rates of the $\Sigma(1430)$ are in good agreement with the Belle results. In addition, in Ref.~\cite{Ren:2026uhb}, the isospin violating $J/\psi\to\Lambda\bar{\Lambda}\pi^0$ reaction was proposed as a means of probing the $\Sigma(1430)$.
  
The BESIII Collaboration recently reported the first observation of the decays $\psi(3686)\to\bar{p}K^+\Sigma^{*0}(1385)$ and $\psi(3686)\to\bar{p}K^{*0}(892)\Sigma^+$, together with their charge-conjugate modes~\cite{BESIII:2026kuj}. The analysis was based on a sample of $(2712.4\pm14.3)\times10^6$ $\psi(3686)$ events. Including the charge-conjugate modes, the measured branching fractions are $(3.94\pm0.15_{\mathrm{stat.}}\pm0.24_{\mathrm{syst.}})\times10^{-5}$ for the former decay and $(4.93\pm0.14_{\mathrm{stat.}}\pm0.30_{\mathrm{syst.}})\times10^{-5}$ for the latter. The measured $\bar{p}K^+$ and $\bar{p}K^{*0}$ invariant mass spectra provide an opportunity to investigate how final-state interactions shape the two near-threshold enhancements.

We first consider the enhancement in the $\bar{p}K^+$ spectrum near the threshold of the charge-conjugate $\bar{K}N$ system. Motivated by the Belle observation and the chiral unitary studies discussed above, we investigate whether the $S$-wave $\bar{K}N$ interaction can account for this enhancement. We denote the near-threshold $\Sigma$ state generated in this framework by $\Sigma(1430)$.

The enhancement in the $\bar{p}K^{*0}$ spectrum lies near the threshold of the $\bar{K}^{*}N$ system. In Ref.~\cite{Oset:2010tof}, the interaction of vector mesons with ground-state octet baryons was studied in a coupled-channel approach. The $I=1$, $S=-1$ sector comprises the $\bar{K}^{*}N$, $\rho\Lambda$, $\rho\Sigma$, $\omega\Sigma$, $K^{*}\Xi$, and $\phi\Sigma$ channels. The calculation yielded two structures in this sector, one of which lies close to the $\bar{K}^{*}N$ threshold. We refer to the state associated with this structure as the $\Sigma(1830)$ and investigate its possible contribution to the observed enhancement.

In the present work, we analyze both $\psi(3686)$ decay modes in a coupled-channel framework. We describe the $\bar{p}K^+$ and $\bar{p}K^{*0}$ final-state interactions using pseudoscalar-meson--baryon and vector-meson--baryon amplitudes, respectively. In the $I=1$ sector, these amplitudes dynamically generate the $\Sigma(1430)$ and $\Sigma(1830)$, respectively. We constrain the model parameters by fitting the available data and compare the calculated invariant mass spectra with the measurements.

The remainder of this paper is organized as follows. Sections~\ref{sec2.A} and~\ref{sec2.B} present the theoretical formalism for $\psi(3686)\to\bar{p}K^{*0}(892)\Sigma^+$ and $\psi(3686)\to\bar{p}K^+\Sigma^{*0}(1385)$, respectively. The numerical results are presented and discussed in Sec.~\ref{sec3}, followed by a summary in Sec.~\ref{sec4}.

\section{Formalism}\label{sec2}

\subsection{The $\psi^\prime\to\bar{p}K^*(892)\Sigma^+$ decay}\label{sec2.A}

\begin{figure}[htbp]
	\centering
	
	%	\centering
	\includegraphics[scale=0.45]{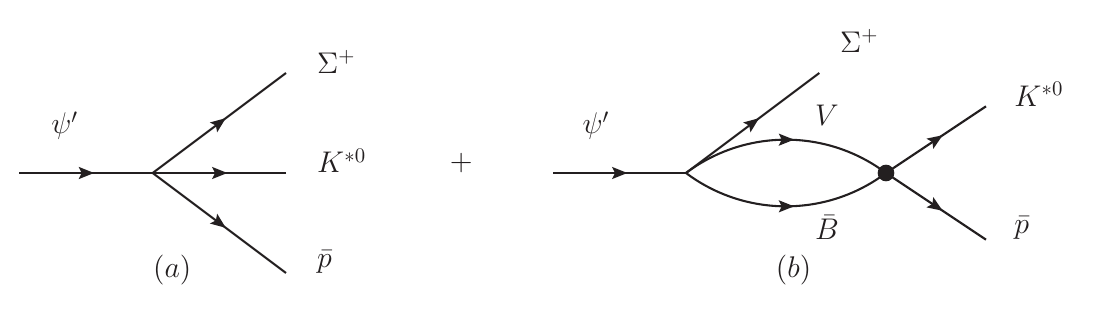}
	
	\caption{Tree-level production and rescattering mechanisms for the decay $\psi^\prime\to\bar pK^{*0}\Sigma^+$.}\label{fig:psiprime_decay_BBV}
\end{figure}

We begin by considering the strong decay $\psi^\prime\to\bar{p}K^*\Sigma^+$.\footnote{Throughout this work, $K^*$ denotes $K^*(892)$, and $\psi^\prime$ denotes $\psi(3686)$.} In this reaction, we account for FSIs in the $K^*\bar{p}$ vector meson--antibaryon subsystem. The relevant hadronic channels are constructed using the vector-meson nonet matrix and the baryon and antibaryon octet matrices, denoted by $V$, $B$, and $\bar B$, respectively.

To identify the meson--baryon and meson--antibaryon channels produced in this decay, we treat $\psi^\prime$ as an SU(3) singlet in the $u$, $d$, $s$ sector. The corresponding flavor structures are represented by the vector-meson nonet matrix and the baryon and antibaryon octet matrices. With ideal mixing in the vector-meson sector, these matrices are given by

\begin{eqnarray}\label{Eq:P_matrix}
	V=
	\left(
	\begin{array}{ccc}
		\frac{1}{\sqrt{2}}\rho^0 + \frac{1}{\sqrt{2}}\omega & \rho^+ & K^{*+} \\[2mm]
		\rho^- & -\frac{1}{\sqrt{2}}\rho^0 + \frac{1}{\sqrt{2}}\omega & K^{*0} \\[2mm]
		K^{*-} & \bar{K}^{*0} & \phi
	\end{array}
	\right),
\end{eqnarray}
\begin{eqnarray}
	\label{Eq:B_matrix}
	B=
	\left(
	\begin{array}{ccc}
		\frac{1}{\sqrt{2}}\Sigma^0 + \frac{1}{\sqrt{6}} \Lambda & \Sigma^+ & p\\[2mm]
		\Sigma^- & -\frac{1}{\sqrt{2}} \Sigma^0 + \frac{1}{\sqrt{6}} \Lambda & n \\[2mm]
		\Xi^- & \Xi^0 & ~-\frac{2}{\sqrt{6}} \Lambda\\
	\end{array}
	\right) ,
\end{eqnarray}
\begin{eqnarray}
	\label{Eq:Bbar_matrix}
	\bar{B}=
	\left(
	\begin{array}{ccc}
		\frac{1}{\sqrt{2}}\bar{\Sigma}^0 + \frac{1}{\sqrt{6}} \bar{\Lambda} & \bar{\Sigma}^+ & \bar{\Xi}^+ \\[2mm]
		\bar{\Sigma}^- & -\frac{1}{\sqrt{2}} \bar{\Sigma}^0 + \frac{1}{\sqrt{6}} \bar{\Lambda }&  \bar{\Xi}^0\\[2mm]
		\bar{p} & \bar{n} & ~-\frac{2}{\sqrt{6}} \bar{\Lambda}\\
	\end{array}
	\right).
\end{eqnarray}

The independent SU(3)-invariant flavor structures can be constructed from single and multiple traces of these matrices: $\langle \bar{B} V B \rangle$, $\langle \bar{B} B V \rangle$, $\langle \bar{B} B\rangle \langle V \rangle$, $\langle \bar{B} V\rangle \langle B \rangle$, $\langle B V\rangle \langle \bar{B} \rangle$, and $\langle \bar{B} \rangle \langle B \rangle \langle V \rangle$. Among these, terms containing $\langle \bar{B} \rangle$ or $\langle B \rangle$ vanish because $B$ and $\bar{B}$ are traceless matrices. The double-trace structure $\langle\bar B B\rangle\langle V\rangle$ is suppressed relative to the single-trace structures in the large-$N_c$ expansion~\cite{Manohar:1998xv,Abreu:2023yvf,Dai:2026zqn,Lyu:2026tdj,CesarSobrinho:2026vhe}. We therefore retain only $\langle B\bar B V\rangle$ and $\langle BV\bar B\rangle$, where the cyclic property of the trace has been used. 

Since we consider rescattering in the vector-meson--antibaryon subsystem, we extract the terms containing the spectator $\Sigma^+$ from the two single-trace structures. This gives
\begin{eqnarray}\label{Eq:meson_baryon_A}
	\langle B\bar{B}V \rangle \Rightarrow
	\Sigma^+\left\{
	\bar{\Sigma}^-\left(\frac{\rho^0}{\sqrt{2}}+\frac{\omega}{\sqrt{2}}\right)
	+\left(-\frac{\bar{\Sigma}^0}{\sqrt{2}}+\frac{\bar{\Lambda}}{\sqrt{6}}\right)\rho^-
	+\bar{\Xi}^0 K^{*-}
	\right\},
\end{eqnarray}
and 
\begin{eqnarray}\label{Eq:meson_baryon_B}
	\langle BV\bar{B} \rangle \Rightarrow
	\Sigma^+\left\{
	\rho^-\left(\frac{\bar{\Sigma}^0}{\sqrt{2}}+\frac{\bar{\Lambda}}{\sqrt{6}}\right)
	+\left(-\frac{\rho^0}{\sqrt{2}}+\frac{\omega}{\sqrt{2}}\right)\bar{\Sigma}^-
	+K^{*0}\bar{p}
	\right\}.
\end{eqnarray}
We assign independent production weights $A$ and $B$ to $\langle B\bar B V\rangle$ and $\langle BV\bar B\rangle$, respectively. The resulting primary production channels and their relative weights are

\begin{equation}\label{Eq:psiprimetoAB}
		\begin{aligned}
			&\psi^\prime \Rightarrow \Sigma^+\left\{
			\frac{A-B}{\sqrt{2}}\bar{\Sigma}^-\rho^0
			+\frac{A+B}{\sqrt{2}}\bar{\Sigma}^-\omega
			+\frac{B-A}{\sqrt{2}}\bar{\Sigma}^0\rho^-
			\right. \\
			&\qquad \left.
			+\frac{A+B}{\sqrt{6}}\bar{\Lambda}\rho^-
			+A\bar{\Xi}^0 K^{*-}
			+BK^{*0}\bar{p}
			\right\}.
		\end{aligned}
\end{equation}

The expressions above determine the flavor coefficients of the channels containing a vector meson, a baryon, and an antibaryon. We now construct the primary production vertex and incorporate the subsequent final-state interactions. We take the spinor structure of the primary vertex to be
\begin{equation}
	t=\bar{u}(p)v(p')\epsilon^\mu\epsilon^\prime_\mu.
\end{equation}
where $\epsilon^\mu$ and $\epsilon^{\prime\mu}$ are the polarization vectors of the final $K^{*0}$ and the initial $\psi^\prime$, respectively, and $p$, $p^\prime$ the momenta of the baryon and antibaryon, respectively. Retaining terms up to first order in the momenta, we use the spinors
\begin{equation}
	u_B(p) = \begin{pmatrix}
		\chi_r  \\[8pt]
		\frac{\vec{\sigma}\cdot\vec{p}}{2m}\chi_r  
	\end{pmatrix},~~~~~~v_{\bar{B}}(p^\prime) = \begin{pmatrix}
		\frac{\vec{\sigma}\cdot\vec{p^\prime}}{2m^\prime}\chi_r^\prime  \\[8pt]
		\chi_r^\prime  
	\end{pmatrix}.
\end{equation}
Using this vertex, we construct the transition amplitude for the $\psi^\prime\to\bar{p}K^*(892)\Sigma^+$ decay from the mechanisms shown in Fig.~\ref{fig:psiprime_decay_BBV}. In the following expressions, all spatial components are treated as contravariant, and repeated spatial indices are summed over. The spinor structure of the tree-level vertex, shown in the first diagram of Fig.~\ref{fig:psiprime_decay_BBV}, reduces to
\begin{align}
	\bar{u}(p)v(p')\epsilon^\mu\epsilon^\prime_\mu \to\epsilon^\mu\epsilon^\prime_\mu\left\{ 
	\frac{\sigma_i p_{\bar p}^{\,i}}{2m^\prime} - \frac{\sigma_i p_{\Sigma^+}^{\,i}}{2m}\right\}.
\end{align}
Both terms contribute to the tree-level amplitude. In the rescattering diagrams, however, terms proportional to the momentum of the intermediate baryon in the loop vanish because the scattering amplitudes are evaluated in the $S$-wave~\cite{Ikeno:2026zpv,Lyu:2026rsm,Ren:2026uhb}. The total transition amplitude for the mechanisms shown in Fig.~\ref{fig:psiprime_decay_BBV} can therefore be written as
\begin{equation}
	\begin{aligned}
		t_{\mathrm{total}}
		=
		\left(
		t_1\frac{\vec\sigma\cdot\vec p_{\bar p}}{2m_{\bar p}}
		-t_2\frac{\vec\sigma\cdot\vec p_{\Sigma^+}}{2m_{\Sigma^+}}
		\right)\epsilon^j\epsilon^\prime_j.
	\end{aligned}
\end{equation}
where 
\begin{equation}\label{Eq:t1_VB}
	t_1=B_t,
\end{equation}
and
\begin{equation}\label{Eq:t2_VB}
\begin{aligned}
	t_2 ={}& B_t
	+A\,
	G_{K^{*-}\bar{\Xi}^0}\,
	t_{K^{*+}\Xi^0,\,\bar{K}^{*0}p}
	\left(M_{K^{*0}\bar p}\right)
	\\
	&+\frac{1}{\sqrt{6}}\left(A+B\right)
	G_{\rho^-\bar{\Lambda}}\,
	t_{\rho^+\Lambda,\,\bar{K}^{*0}p}
	\left(M_{K^{*0}\bar p}\right)
	\\
	&+\frac{1}{\sqrt{2}}\left(B-A\right)
	G_{\rho^-\bar{\Sigma}^0}\,
	t_{\rho^+\Sigma^0,\,\bar{K}^{*0}p}
	\left(M_{K^{*0}\bar p}\right)
	\\
	&+\frac{1}{\sqrt{2}}\left(A+B\right)
	G_{\omega\bar{\Sigma}^-}\,
	t_{\omega\Sigma^+,\,\bar{K}^{*0}p}
	\left(M_{K^{*0}\bar p}\right)
	\\
	&+\frac{1}{\sqrt{2}}\left(A-B\right)
	G_{\rho^0\bar{\Sigma}^-}\,
	t_{\rho^0\Sigma^+,\,\bar{K}^{*0}p}
	\left(M_{K^{*0}\bar p}\right)
	\\
	&+B\,
	G_{K^{*0}\bar{p}}\,
	t_{\bar{K}^{*0}p,\,\bar{K}^{*0}p}
	\left(M_{K^{*0}\bar p}\right).
\end{aligned}
\end{equation}
Here, $B_t$ denotes the effective tree-level production strength, which is treated independently of the rescattering production weights. In Eq.~\eqref{Eq:t2_VB}, we have used the relation $t_{M\bar{B},M\bar{B}}=t_{MB,MB}$ to express the meson--antibaryon amplitudes in terms of the available meson--baryon amplitudes. Since $\Sigma^+$ has isospin $I=1$ and $\psi^\prime$ $I=0$, the coupled terms in Eq.~\eqref{Eq:t2_VB} will have $I=1$. Using the isospin multiplets $(\bar{K}^{*0}, -K^{*-})$, $(\Xi^0, -\Xi^-)$, $(-\rho^+, \rho^0, \rho^-)$, and $(-\Sigma^+, \Sigma^0, \Sigma^-)$, we obtain
\begin{equation}\label{Eq:t2_VB_isospin}
\begin{aligned}
	t_2 ={}& B_t
	+A\,
	G_{K^{*-}\bar{\Xi}^0}\,
	t_{K^*\Xi,\,\bar{K}^*N}^{(I=1)}
	\left(M_{K^{*0}\bar p}\right)
	\\
	&-\frac{1}{\sqrt{6}}\left(A+B\right)
	G_{\rho\bar{\Lambda}}\,
	t_{\rho\Lambda,\,\bar{K}^*N}^{(I=1)}
	\left(M_{K^{*0}\bar p}\right)
	\\
	&+\left(A-B\right)
	G_{\rho\bar{\Sigma}}\,
	t_{\rho\Sigma,\,\bar{K}^*N}^{(I=1)}
	\left(M_{K^{*0}\bar p}\right)
	\\
	&-\frac{1}{\sqrt{2}}\left(A+B\right)
	G_{\omega\bar{\Sigma}^-}\,
	t_{\omega\Sigma,\,\bar{K}^*N}^{(I=1)}
	\left(M_{K^{*0}\bar p}\right)
	\\
	&+B\,
	G_{K^{*0}\bar{p}}\,
	t_{\bar{K}^*N,\,\bar{K}^*N}^{(I=1)}
	\left(M_{K^{*0}\bar p}\right).
\end{aligned}
\end{equation}

For the scattering amplitudes $t_{i,\bar{K}^*N}$, we employ a coupled-channels unitary approach~\cite{Oset:2010tof}, with the following coupled channels: $\bar{K}^{*}N$ (1), $\rho\Lambda$ (2), $\rho\Sigma$ (3), $\omega\Sigma$ (4), $K^{*}\Xi$ (5), and $\phi\Sigma$ (6). The scattering matrix is obtained from
\begin{equation}\label{Eq:BS}
	T = [1-VG]^{-1}V,
\end{equation}
with $G$ being the loop function of a vector meson and a baryon. We evaluate these functions in dimensional regularization using the expression given in Ref.~\cite{Oller:2000fj}, with a regularization scale $\mu=630$~MeV and subtraction constants $a_l(\mu)=-2$. The finite widths of the $\rho$ and $K^*$ mesons are incorporated by convoluting the loop functions with the corresponding meson spectral functions. The loop function $G$ is thus replaced by
~\cite{Oset:2010tof}
\begin{equation}
\begin{aligned}
	\hat{G}(s)=& \frac{1}{N}\int^{(m+2\Gamma_i)^2}_{(m-2\Gamma_i)^2}d\hat{m}^2
	\left(-\frac{1}{\pi}\right) 
	{\rm Im}\,\frac{1}{\hat{m}^2-m^2+{\rm i} \hat{m} \Gamma(\hat{m})} \\
	&\times G(s,\hat{m}^2,m^2_B)
	\label{Gconvolution}
\end{aligned}
\end{equation}
with
\begin{equation}
	N=\int^{(m+2\Gamma_i)^2}_{(m-2\Gamma_i)^2}d\hat{m}^2
	\left(-\frac{1}{\pi}\right){\rm Im}\,\frac{1}{\hat{m}^2-m^2+{\rm i} \hat{m} \Gamma(\hat{m})}
	\label{Norm}
\end{equation}
being the normalization factor, and $\Gamma_i$ the decay width of the meson ($i=\rho,K^*$), which we take to be 149.4 MeV and 50.5 MeV for the $\rho$ and  $K^*$ meson, respectively. The energy dependent width $\Gamma(\hat{m})$ for the $\rho$ meson, obtained from its decay into two pions in $p$-wave, is given by
\begin{equation}
	\Gamma(\hat{m})=\Gamma_\rho \frac{m_\rho^2}{\hat{m}^2}
	\left(\frac{\hat{m}^2-4m^2_\pi}{m^2_\rho-4m^2_\pi}\right)^{3/2}
	\theta(\hat{m}-2m_\pi) \ .
	\label{gamma}
\end{equation}
A similar expression gives the energy dependent width of the $K^*$ meson from its decay into a $K$ meson and a pion. For the potential $V$ in Eq.~\eqref{Eq:BS}, we have 
\begin{equation}
V_{ij}=-C_{ij}\frac{1}{4f^{2}}\left(k^{0}+k^{\prime0}\right)\vec{\epsilon}\cdot\vec{\epsilon}^{~\prime}
\end{equation}
where $k^{0}$ and $k^{\prime0}$ are the energies of the incoming and outgoing vector mesons, respectively, and $f=93$~MeV. The coefficients $C_{ij}$ for the $I=1$ sector are listed in Table~\ref{Tab:Cij}.

\begin{table}
	\centering
	\caption{Coefficients $C_{ij}$ of the vector-meson--baryon interaction in the $I=1$ sector.~\cite{Oset:2010tof}.}
	\label{Tab:Cij}
	$
	\begin{array}{c|cccccc}
		\hline
		& \bar{K}^{*}N & \rho\Lambda & \rho\Sigma & \omega\Sigma & K^{*}\Xi & \phi\Sigma \\
		\hline
		\bar{K}^{*}N & 1 & -\sqrt{\frac{3}{2}} & -1 & -\sqrt{\frac{3}{2}}\frac{1}{\sqrt{3}} & 0 & -\sqrt{\frac{3}{2}}\left(-\sqrt{\frac{2}{3}}\right) \\
		\rho\Lambda & & 0 & 0 & 0 & -\sqrt{\frac{3}{2}} & 0 \\
		\rho\Sigma & & & 2 & 0 & 1 & 0 \\
		\omega\Sigma & & & & 0 & -\sqrt{\frac{3}{2}}\frac{1}{\sqrt{3}} & 0 \\
		K^{*}\Xi & & & & & 1 & -\sqrt{\frac{3}{2}}\left(-\sqrt{\frac{2}{3}}\right) \\
		\phi\Sigma & & & & & & 0 \\
		\hline
	\end{array}
	$
\end{table}

By summing and averaging $|t_{\text{Total}}|^2$ over the particle spins, we obtain
\begin{equation}
	\begin{aligned}
	\overline{\sum}|t_{\text{Total}}|^2=&2\left\{
		|t_1|^2\frac{\vec{p}_{\bar{p}}^{\,2}}{4m_{\bar{p}}^2}
		+|t_2|^2\frac{\vec{p}_{\Sigma^+}^{\,2}}{4m_{\Sigma^+}^2}
		-2\operatorname{Re}\left(t_1t_2^*\right)
		\frac{\vec{p}_{\Sigma^+}\cdot\vec{p}_{\bar{p}}}{2m_{\Sigma^+}\,2m_{\bar{p}}}
		\right\},
	\end{aligned}
\end{equation}
where we applied the polarization sum $\sum\epsilon_{i}\epsilon_{j}=\delta_{ij}$ and $\sigma_i\sigma_j=\delta_{ij}+i\epsilon_{ijk}\sigma_{k}$. The $\Sigma^+$ and $\bar{p}$ three-momenta are evaluated in the rest frame of the parent meson, with magnitudes
\begin{equation}
	|\vec{p}_{\Sigma^+}|=\frac{\lambda^{1/2}\!\left(M_{\psi^\prime}^2,~M_{\Sigma^+}^2,~M_{K^{*0}\bar p}^2\right)}{2M_{\psi^\prime}},
\end{equation}
and
\begin{equation}
|\vec{p}_{\bar p}|=\frac{\lambda^{1/2}\!\left(M_{\psi^\prime}^2,~M_{\bar p}^2,~M_{K^{*0}\Sigma^+}^2\right)}{2M_{\psi^\prime}},
\end{equation}
where the Källén function is defined as
\begin{equation}
	\lambda(x,y,z)=x^2+y^2+z^2-2xy-2xz-2yz.
\end{equation}
The scalar product of the two three-momenta can be expressed in terms of the antiproton--hyperon invariant mass as
\begin{equation}
    \vec{p}_{\bar p}\cdot\vec{p}_{\Sigma^+}=\frac{1}{2}\left(M_{\bar p}^2+M_{\Sigma^+}^2+2E_{\bar p}E_{\Sigma^+}-M_{\Sigma^+\bar p}^2\right),
\end{equation}
where the antiproton and hyperon energies are given by
\begin{equation}
	E_{\bar p}=\frac{M_{\psi^\prime}^2+M_{\bar p}^2-M_{K^{*0}\Sigma^+}^2}{2M_{\psi^\prime}},
\end{equation}
and
\begin{equation}
	E_{\Sigma^+}=\frac{M_{\psi^\prime}^2+M_{\Sigma^+}^2-M_{K^{*0}\bar p}^2}{2M_{\psi^\prime}}.
\end{equation}
The invariant masses appearing in these expressions correspond to the final-state subsystems indicated by their subscripts.

Using the field-normalization conventions of Mandl and Shaw~\cite{Mandl:1985bg}, we combine the spin-summed and polarization-averaged squared amplitude with the three-body phase space to obtain
\begin{equation}
	\dfrac{d^2\Gamma}{dM_{12}dM_{23}} = \frac{1}{(2\pi)^3}\dfrac{2M_{\bar{p}}~2M_{\Sigma^+}}{32M_{\psi^\prime}^3}\overline{\sum}\sum|t_{\text{Total}}|^2~2M_{12}~2M_{23}.
\end{equation}

It is worth noting that final-state interactions in the $K^*\Sigma$ channel may also contribute to the present process. In Ref.~\cite{Oset:2010tof}, these interactions were investigated within a coupled-channel framework involving $\rho N$, $\omega N$, $\phi N$, $K^*\Lambda$, and $K^*\Sigma$. Two peaks were found at approximately 1700 and 1970~MeV, both well below the $K^*\Sigma$ threshold of approximately 2085~MeV. We therefore neglect the $K^*\Sigma$ final-state interactions in the present calculation.

\subsection{The $\psi^\prime\to\bar{p}K^+\Sigma^*(1385)$ decay}\label{sec2.B}

\begin{figure}[htbp]
	\centering
	
	%	\centering
	\includegraphics[scale=0.45]{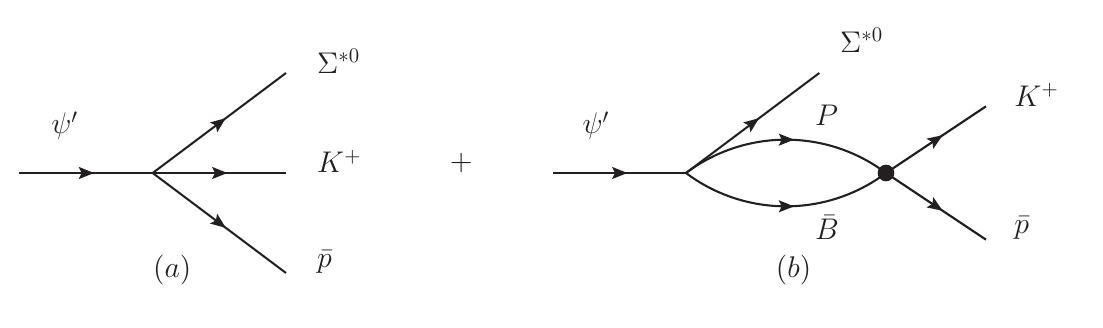}
	
	\caption{Tree-level production and rescattering mechanisms for the decay $\psi^\prime\to\bar pK^+\Sigma^{*0}$.
	}\label{fig:psiprime_decay_BBstarP}
\end{figure}

We now turn to the decay $\psi^\prime\to\bar pK^+\Sigma^{*0}(1385)$. The coupled-channel calculation of Ref.~\cite{Sarkar:2004jh} generates the $N^*(1520)$ resonance and a weak peak just below 1900~MeV, as shown in the upper panel of Fig.~5 therein. The latter lies near the $K\Sigma^*$ threshold at approximately 1878~MeV and has only a small strength. The associated enhancement becomes negligible above 1900~MeV. Moreover, the experimental mass distribution shown in Fig.~2(c) of Ref.~\cite{BESIII:2026kuj} exhibits no clear enhancement near the $K\Sigma^*$ threshold. We therefore neglect the final-state interactions in the $K\Sigma^*$ channel and its coupled channels, including $\pi\Delta$, and focus instead on the $\bar{p} K^+$ channel and its coupled channels. 

Since the final state considered here contains a decuplet baryon, we determine the hadronization weights using SU(3) Clebsch–Gordan coefficients~\cite{McNamee:1964xq} rather than the SU(3) trace structures involving $B$, $\bar B$, and $V$ used in the preceding subsection. Given that $\Sigma^{*0}$ has isospin $I=1$ and $\psi^\prime$ is an isosinglet, the remaining meson--antibaryon subsystem must have $I=1$ and $S=+1$. The corresponding charge channels are $K^+\bar p$, $K^0\bar n$, $\pi^+\bar{\Sigma}^-$, $\pi^0\bar{\Sigma}^0$, $\pi^-\bar{\Sigma}^+$, $\pi^0\bar{\Lambda}$, $\eta\bar{\Sigma}^0$, $\bar K^0\bar{\Xi}^0$, and $K^-\bar{\Xi}^+$. Using these coefficients, we obtain

\begin{equation}\label{Eq:psiprimetoBstarBP}
	\begin{aligned}
		&\psi^\prime \Rightarrow\Sigma^{*0}\left\{
		-\frac{1}{\sqrt{15}}K^+\bar p
		+\frac{1}{\sqrt{15}}K^0\bar n
		+\frac{1}{\sqrt{15}}\pi^+\bar{\Sigma}^-
		+0\,\pi^0\bar{\Sigma}^0
		\right. \\
		&\left.-\frac{1}{\sqrt{15}}\pi^-\bar{\Sigma}^+
		+\frac{1}{\sqrt{5}}\pi^0\bar{\Lambda}
		-\frac{1}{\sqrt{5}}\eta\bar{\Sigma}^0
		-\frac{1}{\sqrt{15}}\bar K^0\bar{\Xi}^0
		+\frac{1}{\sqrt{15}}K^-\bar{\Xi}^+
		\right\}.
	\end{aligned}
\end{equation}

Equation~\eqref{Eq:psiprimetoBstarBP} specifies the relative production weights of the meson--antibaryon channels, but the spin and momentum dependence of the primary production vertex must still be determined. The initial $\psi^\prime$ has spin-parity $1^-$, whereas the final antibaryon, pseudoscalar meson, and decuplet baryon have spin-parities $1/2^-$, $0^-$, and $3/2^+$, respectively. The product of their intrinsic parities is positive. Parity conservation therefore requires the sum of the two relative orbital angular momenta in the three-body final state to be odd. The lowest-order production vertex consequently contains one power of a final-state momentum.

The transition between the spin-$1/2$ antibaryon and the spin-$3/2$ decuplet baryon is described by the spin-transition operator $\vec S^{\,\dagger}$. Within the non-relativistic spin treatment adopted here, a suitable scalar structure is
\begin{equation}
\left(\vec S^{\,\dagger}\times\vec{\tilde{p}}_{K^+}\right)\cdot\vec\epsilon,
\end{equation}
where $\vec\epsilon$ is the polarization vector of the initial $\psi^\prime$ and $\vec{\tilde{p}}_{K^+}$ is the kaon momentum in the $\Sigma^{*0}$ rest frame. The tilde will be used to distinguish momenta in this frame from those in the $\psi^\prime$ rest frame. We neglect relativistic corrections to the spin operators and use the momentum transformation below to examine the structures generated by the rescattering loops.

The kaon momentum in the $\Sigma^{*0}$ rest frame is obtained by boosting from the $\psi^\prime$ rest frame~\cite{Crdoba1995ProjectileDE,Lyu:2026rsm}
\begin{equation}
\vec{\tilde{p}}_{K^+}
=
\vec p_{K^+}
+
\left[
\left(\frac{E_{\Sigma^{*0}}}{M_{\Sigma^{*0}}}-1\right)
\frac{\vec p_{K^+}\cdot\vec p_{\Sigma^{*0}}}
{\vec p_{\Sigma^{*0}}^{\,2}}
-\frac{\omega_{K^+}}{M_{\Sigma^{*0}}}
\right]\vec p_{\Sigma^{*0}},
\end{equation}
where
\begin{equation}
E_{\Sigma^{*0}}
=\sqrt{M_{\Sigma^{*0}}^2+\vec p_{\Sigma^{*0}}^{\,2}},
\qquad
\omega_{K^+}
=\sqrt{m_{K^+}^2+\vec p_{K^+}^{\,2}}.
\end{equation}
All untilded three-momenta in this expression are defined in the $\psi^\prime$ rest frame. The expression at $\vec p_{\Sigma^{*0}}=0$ is understood through its continuous limit.

The boost separates the kaon momentum into an external-kaon term and a term along the spectator momentum:
\begin{equation}
\vec{\tilde{p}}_{K^+}
=\vec p_{K^+}+\alpha_{K^+}\vec p_{\Sigma^{*0}},
\end{equation}
where $\alpha_{K^+}$ is specified by the boost formula above. We next consider how these two terms contribute when an intermediate meson--antibaryon channel $P_i\bar B_i$ rescatters into $K^+\bar p$.

The total four-momentum of the interacting pair is
\begin{equation}
P^\mu=p_{\psi^\prime}^\mu-p_{\Sigma^{*0}}^\mu
=p_{K^+}^\mu+p_{\bar p}^\mu,
\end{equation}
so that $\vec P=-\vec p_{\Sigma^{*0}}$ in the $\psi^\prime$ rest frame. In a rescattering loop, the kaon momentum is replaced by the internal momentum $\vec q$ of the intermediate pseudoscalar meson. With the meson--antibaryon interaction retained in $S$-wave, the momentum-dependent loop integral has the form~\cite{Duan:2024ygq}
\begin{equation}
\vec I_i(P)
=i\int\frac{d^4q}{(2\pi)^4}\,
\vec q\,\mathcal K_i(P,q)
=\mathcal F_i\vec P.
\end{equation}
Here, $\mathcal K_i(P,q)$ is the scalar loop kernel for channel $i$, and $\mathcal F_i$ is a scalar function. The last equality follows from rotational symmetry: after integration over the internal momentum, $\vec P$ is the only external vector available. The integral vanishes when $\vec P=0$, but it need not vanish for a moving meson--antibaryon pair.

Consequently, the intermediate-meson-momentum term produces
\begin{equation}
\left(\vec S^{\,\dagger}\times\vec I_i(P)\right)\cdot\vec\epsilon
=\tilde{A}
\left(\vec S^{\,\dagger}\times\vec p_{\Sigma^{*0}}\right)\cdot\vec\epsilon.
\end{equation}
The second term in the boosted momentum is already proportional to $\vec p_{\Sigma^{*0}}$ and generates the same spin--momentum structure after integration. Thus, both parts of the boosted momentum contribute to a loop amplitude of the form
\begin{equation}
t_{\mathrm{loop}}
=\tilde{A}
\left(\vec S^{\,\dagger}\times\vec p_{\Sigma^{*0}}\right)\cdot\vec\epsilon~\tilde{t}_{\mathrm{loop}}.
\end{equation}
Here, $\tilde A$ is an effective production strength, including the overall normalization of the hadronization weights, while $\tilde t_{\mathrm{loop}}$ contains the relative channel weights and the loop and rescattering contributions:
\begin{equation}
	\begin{aligned}
		\tilde{t}_{\mathrm{loop}}
		&=
		-\tilde{G}_{\bar{K}N}\,t_{K^-p,\,K^-p}
		+\tilde{G}_{\bar{K}N}\,t_{\bar{K}^0n,\,K^-p}\\
		&\qquad
		+\tilde{G}_{\pi\Sigma}\,t_{\pi^-\Sigma^+,\,K^-p}
		-\tilde{G}_{\pi\Sigma}\,t_{\pi^+\Sigma^-,\,K^-p}\\
		&\qquad
		+\sqrt{3}\,\tilde{G}_{\pi\Lambda}\,t_{\pi^0\Lambda,\,K^-p}
		-\sqrt{3}\,\tilde{G}_{\eta\Sigma^0}\,t_{\eta\Sigma^0,\,K^-p}\\
		&\qquad
		-\tilde{G}_{K\Xi}\,t_{K^0\Xi^0,\,K^-p}
		+\tilde{G}_{K\Xi}\,t_{K^+\Xi^-,\,K^-p}.
	\end{aligned}
\end{equation}
For the scattering amplitudes $t_{i,K^-p}$, we employ the chiral unitary approach described in Ref.~\cite{Oset:1997it}, with the following coupled channels: $K^-p$ (1), $\bar{K}^{0}n$ (2), $\pi^{0}\Lambda$ (3), $\pi^{0}\Sigma^{0}$ (4), $\eta\Lambda$ (5), $\eta\Sigma^{0}$ (6), $\pi^{+}\Sigma^{-}$ (7), $\pi^{-}\Sigma^{+}$ (8), $K^{+}\Xi^{-}$ (9), and $K^{0}\Xi^{0}$ (10). We evaluate the loop functions $\tilde{G}$ using a three--momentum cutoff method with $q_{\text{max}}=630$~MeV.

At tree level, however, the kaon momentum is an external momentum and is not integrated over in the amplitude. The two momentum structures therefore remain distinct. We parametrize the direct production amplitude as
\begin{equation}
t_{\mathrm{tree}}
=\tilde{C}\left(\vec S^{\,\dagger}\times\vec p_{\Sigma^{*0}}\right)\cdot\vec\epsilon
+\tilde{B}\left(\vec S^{\,\dagger}\times\vec p_{K^+}\right)\cdot\vec\epsilon,
\end{equation}
where $\tilde{B}$ and $\tilde{C}$ are independent effective production parameters. The direct-channel weight $-1/\sqrt{15}$ has been absorbed into their definitions. All three-momenta in this expression are evaluated in the $\psi^\prime$ rest frame.

Adding the tree-level and loop contributions, the full amplitude can be expressed in terms of the same two momentum structures:
\begin{equation}
\begin{aligned}
	\tilde{t}_{\text{total}}
	&=t_{\mathrm{tree}}+t_{\mathrm{loop}}\\
	&=\left(\vec S^{\,\dagger}\times\vec p_{\Sigma^{*0}}\right)\cdot\vec\epsilon\left\{\tilde{C}-\tilde{A}~\tilde{t}_{\mathrm{loop}}\right\}
	+\tilde{B}\left(\vec S^{\,\dagger}\times\vec p_{K^+}\right)\cdot\vec\epsilon\\
	&=\vec S^{\,\dagger}\times\left[\vec p_{\Sigma^{*0}}\left(\tilde{C}-\tilde{A}~\tilde{t}_{\mathrm{loop}}\right)+\tilde{B}~\vec p_{K^+}\right]\vec\epsilon,
\end{aligned}
\end{equation}
Thus, the spectator-momentum structure receives contributions from both direct production and rescattering, whereas the external-kaon-momentum structure appears explicitly only in the tree-level amplitude.

To evaluate the squared amplitude, we sum over the spins of the final baryons and average over the three polarization states of the initial $\psi^\prime$. The spin-transition operator is normalized in the spherical basis according to
\begin{equation}
	\left\langle\frac{3}{2}M\left|S^\dagger_\nu\right|\frac{1}{2}m\right\rangle=C\left(\frac{1}{2},\mathbf{1},\frac{3}{2}; m,\nu,M\right)\left\langle\frac{3}{2}\left|\left|S^\dagger\right|\right|\frac{1}{2}\right\rangle,
\end{equation}
where $\nu$ is the spherical component of the rank-$\mathbf{1}$ operator $\vec{S}^\dagger$, which makes explicit use of the Wigner-Eckart theorem with Clebsch–Gordan coefficients, and the reduced matrix element is defined as unity~\cite{Ericson:1988gk,Oset:1981ih,Duan:2024czu,Lyu:2026ack}. In Cartesian coordinates, the corresponding identity is
\begin{equation}
S_iS_j^\dagger=\frac{2}{3}\delta_{ij}-\frac{i}{3}\epsilon_{ijk}\sigma_k,
\end{equation}
where $\sigma_k$ denotes a Pauli matrix. Using the completeness of the spin-$3/2$ states, the final-state spin sum gives
\begin{equation}
\begin{aligned}
	&\sum_{M,m}
	\langle M|S_i^\dagger|m\rangle
	\langle m|S_{i'}|M\rangle\\
	&\qquad
	=\sum_m\langle m|S_{i'}S_i^\dagger|m\rangle\\
	&\qquad
	=\frac23\delta_{ii'}\sum_m1
	-\frac{i}{3}\epsilon_{i'ik}\sum_m\langle m|\sigma_k|m\rangle
	=\frac43\delta_{ii'}.
\end{aligned}
\end{equation}
Here, the sum over the two spin states of the final antibaryon gives $\sum_m1=2$, while the Pauli-matrix term vanishes because $\operatorname{Tr}\sigma_k=0$. In the $\psi^\prime$ rest frame, the polarization sum satisfies
\begin{equation}
\frac13\sum_{\lambda}
\epsilon_k^{(\lambda)}\epsilon_{k'}^{(\lambda)*}
=\frac13\delta_{kk'}.
\end{equation}

Writing the cross product in the total amplitude explicitly in Cartesian components, we obtain
\begin{equation}
\begin{aligned}
	\overline{\sum}\left|\tilde t_{\mathrm{total}}\right|^2
	&=\frac49\epsilon_{ijk}\epsilon_{ij'k}
	\left[
	p_{\Sigma^{*0},j}\left(\tilde C-\tilde A\tilde t_{\mathrm{loop}}\right)
	+\tilde Bp_{K^+,j}
	\right]\\
	&\qquad\times
	\left[
	p_{\Sigma^{*0},j'}\left(\tilde C-\tilde A\tilde t_{\mathrm{loop}}\right)^*
	+\tilde Bp_{K^+,j'}
	\right]\\
	&=\frac89
	\left[
	\vec p_{\Sigma^{*0}}\left(\tilde C-\tilde A\tilde t_{\mathrm{loop}}\right)
	+\tilde B\vec p_{K^+}
	\right]\\
	&\qquad\cdot
	\left[
	\vec p_{\Sigma^{*0}}\left(\tilde C-\tilde A\tilde t_{\mathrm{loop}}\right)^*
	+\tilde B\vec p_{K^+}
	\right],
\end{aligned}
\end{equation}
where repeated Cartesian indices are summed, and the identity $\epsilon_{ijk}\epsilon_{ij'k}=2\delta_{jj'}$ has been used. The bar denotes averaging over the initial polarization states, while the sum includes both final-state baryon spins.

The spin-summed and polarization-averaged squared amplitude is therefore
\begin{equation}
\begin{aligned}
	\overline{\sum}\left|\tilde t_{\mathrm{total}}\right|^2
	=\frac89\Big\{&
	\vec p_{\Sigma^{*0}}^{\,2}
	\left|\tilde C-\tilde A\tilde t_{\mathrm{loop}}\right|^2
	+\tilde B^2\vec p_{K^+}^{\,2}\\
	&+2\vec p_{\Sigma^{*0}}\cdot\vec p_{K^+}
	\operatorname{Re}\left[
	\left(\tilde C-\tilde A\tilde t_{\mathrm{loop}}\right)\tilde B
	\right]
	\Big\}.
\end{aligned}
\end{equation}
The three-momenta entering the squared amplitude are evaluated in the $\psi^\prime$ rest frame. Their magnitudes are given by
\begin{equation}
\begin{aligned}
	|\vec p_{\Sigma^{*0}}|
	&=\frac{\lambda^{1/2}\left(M_{\psi^\prime}^2,M_{\Sigma^{*0}}^2,M_{K^+\bar p}^{\,2}\right)}{2M_{\psi^\prime}},\\
	|\vec p_{K^+}|
	&=\frac{\lambda^{1/2}\left(M_{\psi^\prime}^2,m_{K^+}^2,M_{\Sigma^{*0}\bar p}^{\,2}\right)}{2M_{\psi^\prime}}.
\end{aligned}
\end{equation}
The corresponding energies are
\begin{equation}
\begin{aligned}
	\omega_{K^+}
	&=\frac{M_{\psi^\prime}^2+m_{K^+}^2-M_{\Sigma^{*0}\bar p}^{\,2}}{2M_{\psi^\prime}},\\
	E_{\Sigma^{*0}}
	&=\frac{M_{\psi^\prime}^2+M_{\Sigma^{*0}}^2-M_{K^+\bar p}^{\,2}}{2M_{\psi^\prime}}.
\end{aligned}
\end{equation}
Using the invariant mass of the $K^+\Sigma^{*0}$ pair,
\begin{equation}
M_{K^+\Sigma^{*0}}^{\,2}
=m_{K^+}^2+M_{\Sigma^{*0}}^2
+2\omega_{K^+}E_{\Sigma^{*0}}
-2\vec p_{K^+}\cdot\vec p_{\Sigma^{*0}},
\end{equation}
the momentum scalar product can be written as
\begin{equation}
\vec p_{K^+}\cdot\vec p_{\Sigma^{*0}}
=\frac12\left(
m_{K^+}^2+M_{\Sigma^{*0}}^2
+2\omega_{K^+}E_{\Sigma^{*0}}
-M_{K^+\Sigma^{*0}}^{\,2}
\right).
\end{equation}

Using the Mandl and Shaw normalization for the meson and baryon fields~\cite{Mandl:1985bg}, we obtain the differential decay width by combining the spin-summed squared amplitude:
\begin{equation}
	\dfrac{d^2\Gamma}{dM_{12}dM_{23}} = \frac{1}{(2\pi)^3}\dfrac{2M_{\bar{p}}~2M_{\Sigma^{*0}}}{32M_{\psi^\prime}^3}\overline{\sum}\sum|\tilde t_{\text{Total}}|^2~2M_{12}~2M_{23}.
\end{equation}
The three pair invariant masses satisfy
\begin{equation}
M_{K^+\bar p}^{\,2}
+M_{\Sigma^{*0}\bar p}^{\,2}
+M_{K^+\Sigma^{*0}}^{\,2}
=
M_{\psi^\prime}^2+m_{K^+}^2+M_{\Sigma^{*0}}^2+M_{\bar p}^2.
\end{equation}
Thus, all momentum factors in the squared amplitude can be expressed in terms of the two invariant masses used in the differential decay width.

\section{Results and Discussions}\label{sec3}

\begin{figure}[htbp]
	\centering
	
	\subfigure[]{
		\includegraphics[scale=0.43]{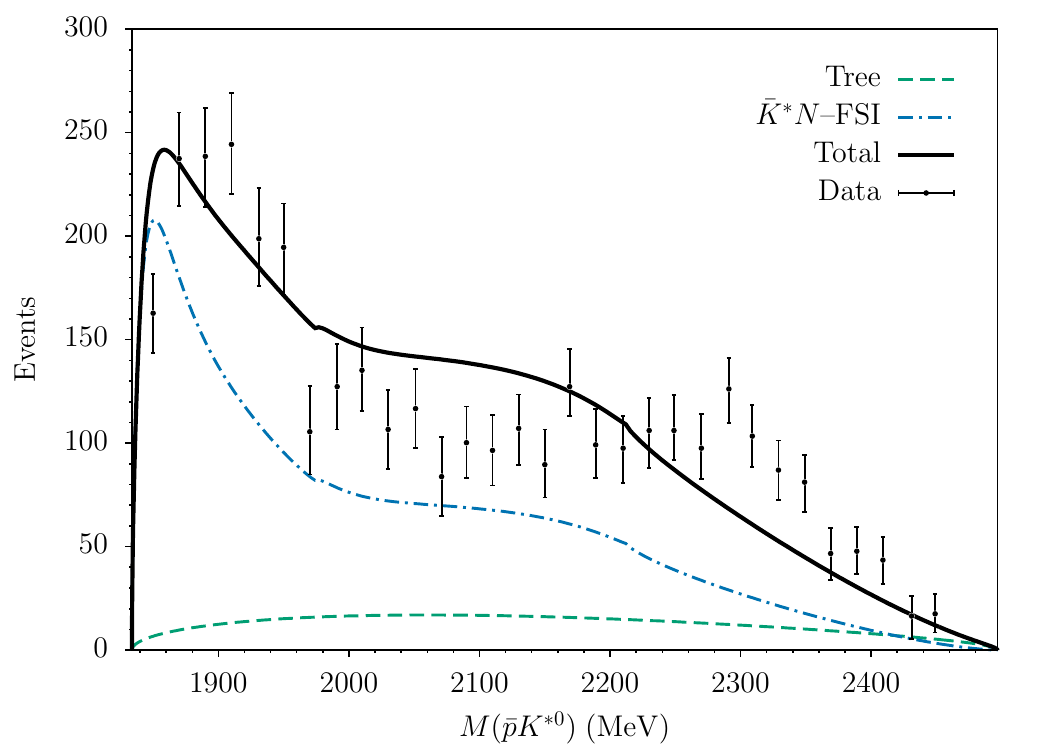}
		\label{fig:fit_pbarKstar}
	}\hfill
	\subfigure[]{
		\includegraphics[scale=0.43]{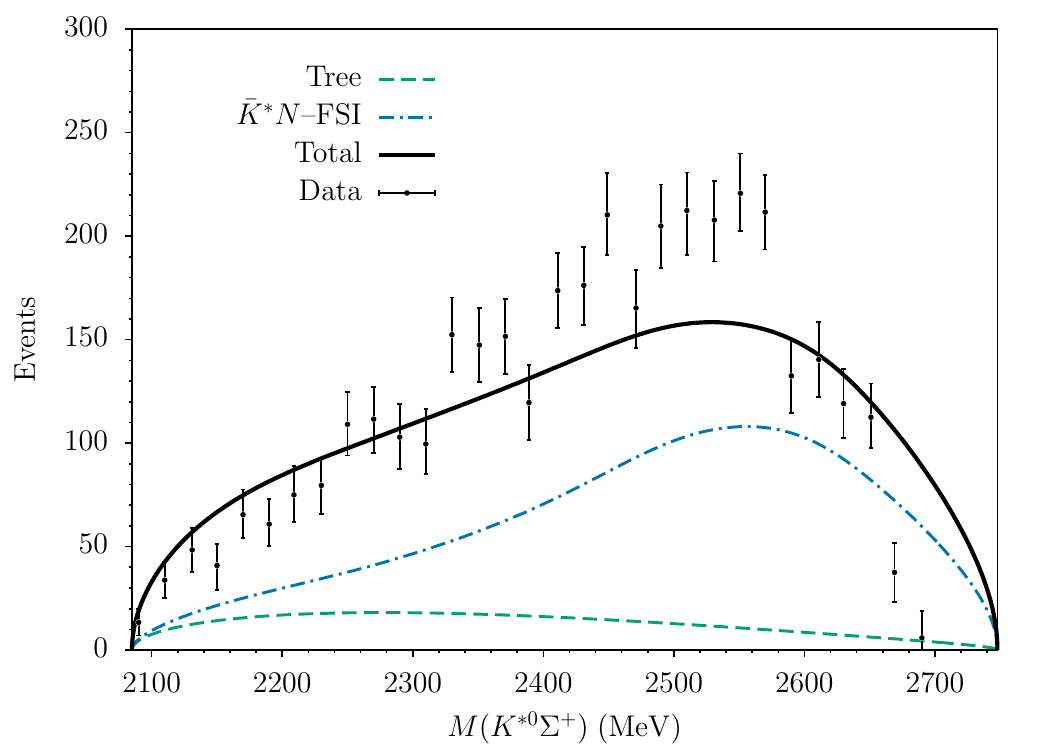}
		\label{fig:fit_KstarSigma}
	}\hfill
	\subfigure[]{
		\includegraphics[scale=0.43]{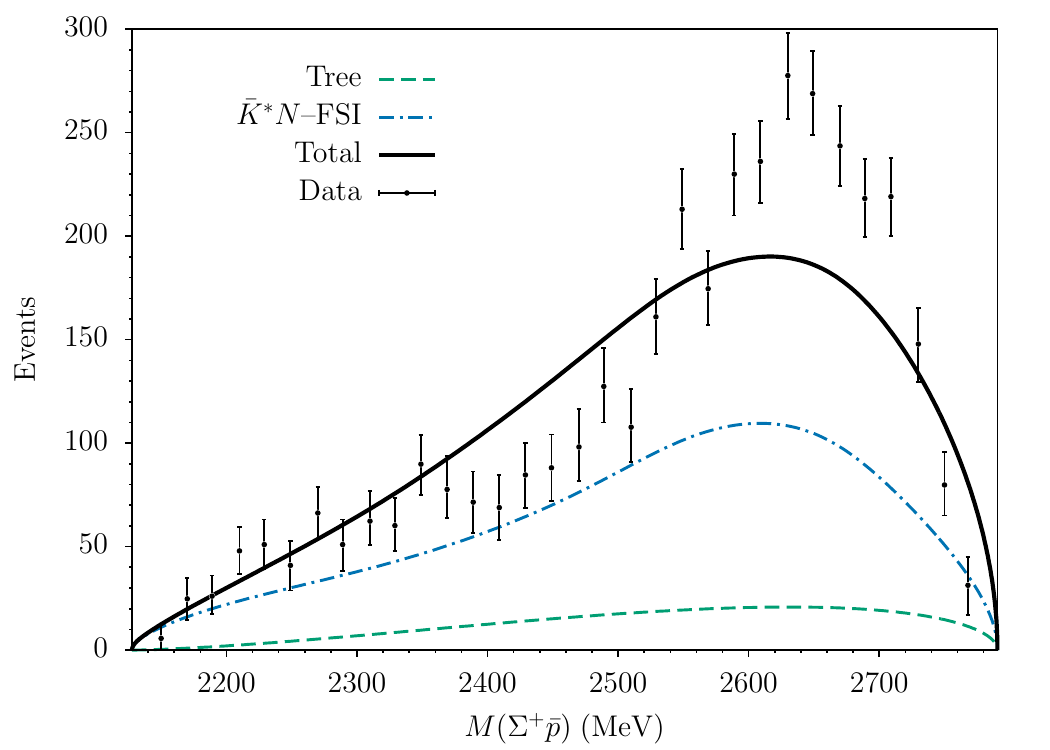}
		\label{fig:fit_SigmapBar}
	}
	\caption{Invariant mass distributions of $\bar{p}K^{*0}$~(a), $K^{*0}\Sigma^{+}$~(b), $\Sigma^{+}\bar{p}$~(c) for the $\psi(3686)\to\bar{p}K^*(892)\Sigma^+$ process. The ``Data" denotes the BESIII experimental data~\cite{BESIII:2026kuj}.}
	\label{fig:invariant_mass_distribution_BBV}
\end{figure}

\begin{figure}[htbp]
	\centering
	
	\subfigure[]{
		\includegraphics[scale=0.43]{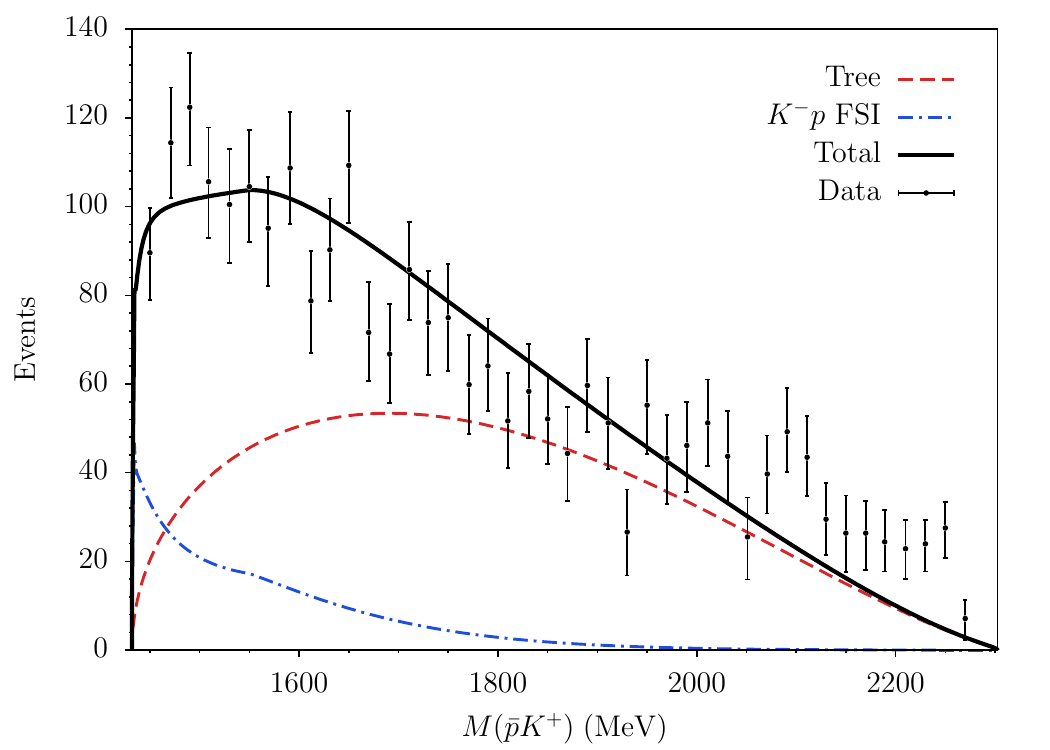}
		\label{fig:fit_pbarK}
	}\hfill
	\subfigure[]{
		\includegraphics[scale=0.43]{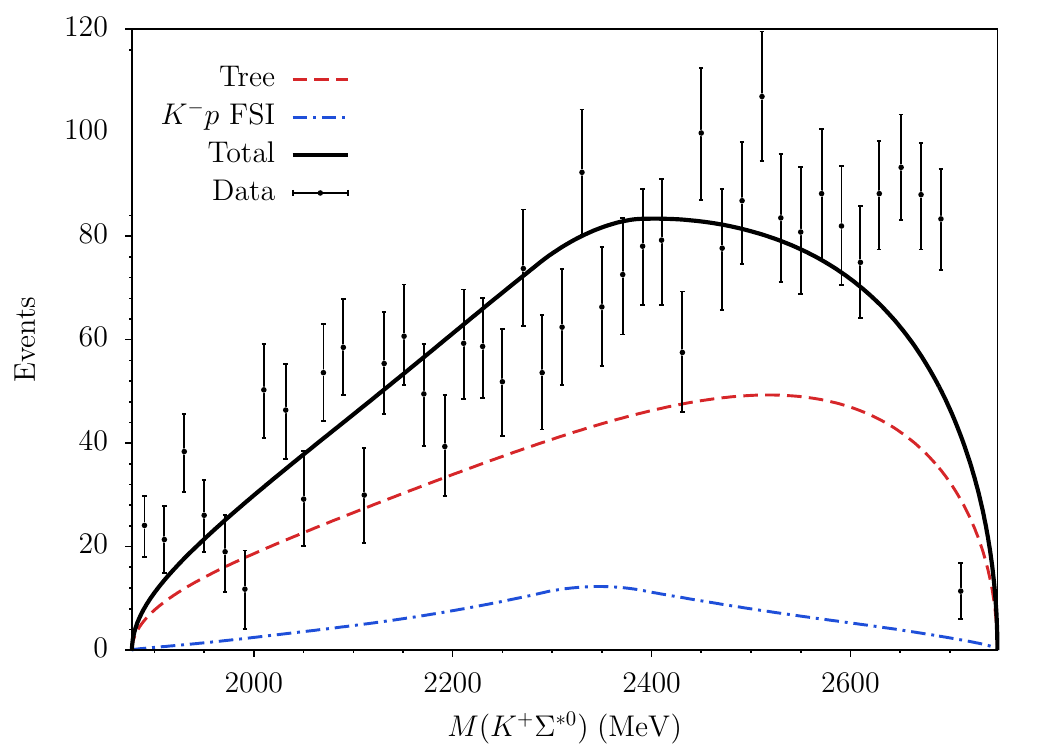}
		\label{fig:fit_KSigmastar}
	}\hfill
	\subfigure[]{
		\includegraphics[scale=0.43]{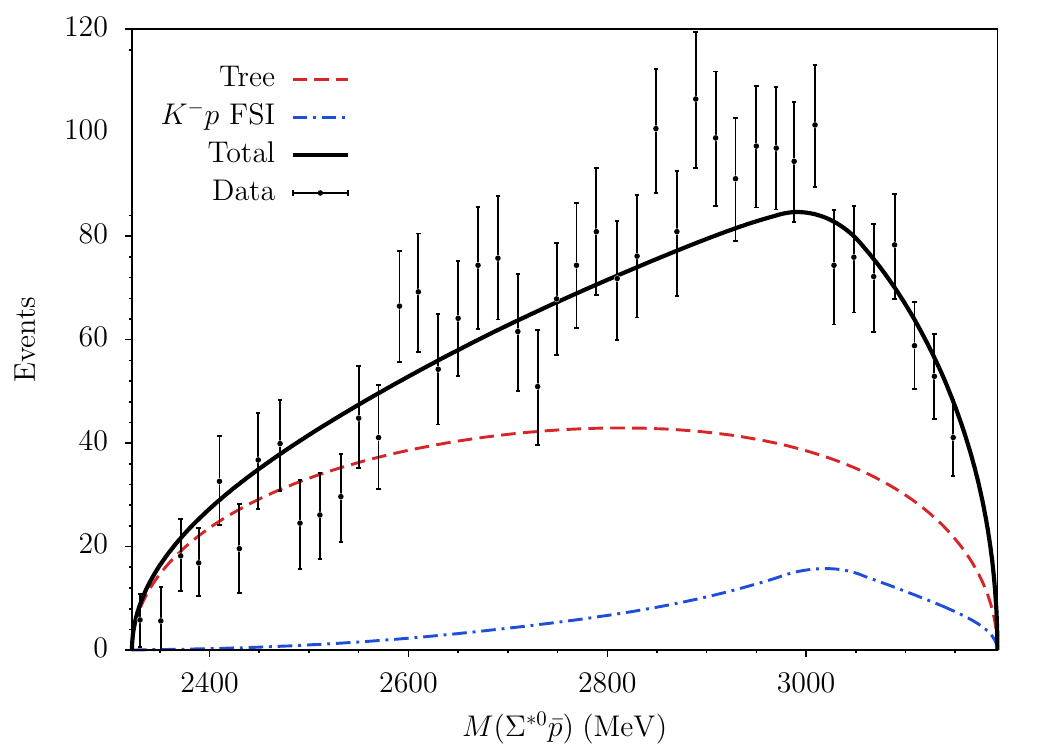}
		\label{fig:fit_SigmastarPbar}
	}
	\caption{Invariant mass distributions of $\bar{p}K^{+}$~(a), $K^{+}\Sigma^{*0}$~(b), $\Sigma^{*0}\bar{p}$~(c) for the $\psi(3686)\to\bar{p}K^+\Sigma^*(1385)$ process. The ``Data" denotes the BESIII experimental data~\cite{BESIII:2026kuj}.}
	\label{fig:invariant_mass_distribution_BBP}
\end{figure}

\begin{figure}[htbp]
	\centering
	
	\subfigure[]{
		\includegraphics[scale=0.43]{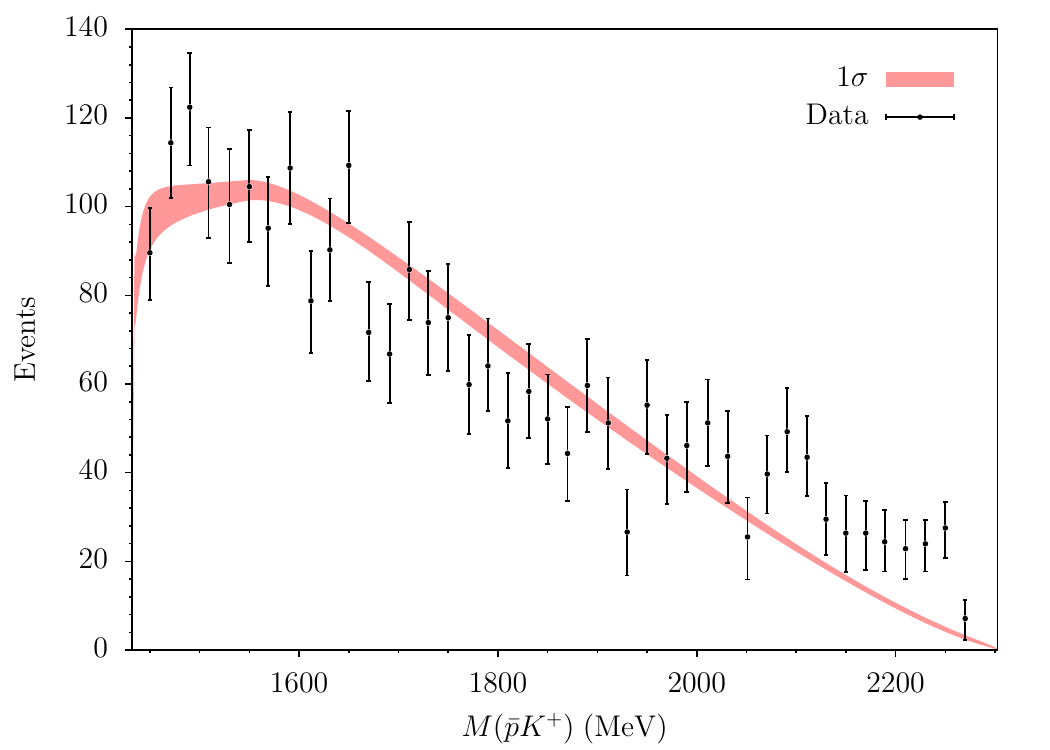}
		\label{fig:error_pbarK}
	}\hfill
	\subfigure[]{
		\includegraphics[scale=0.43]{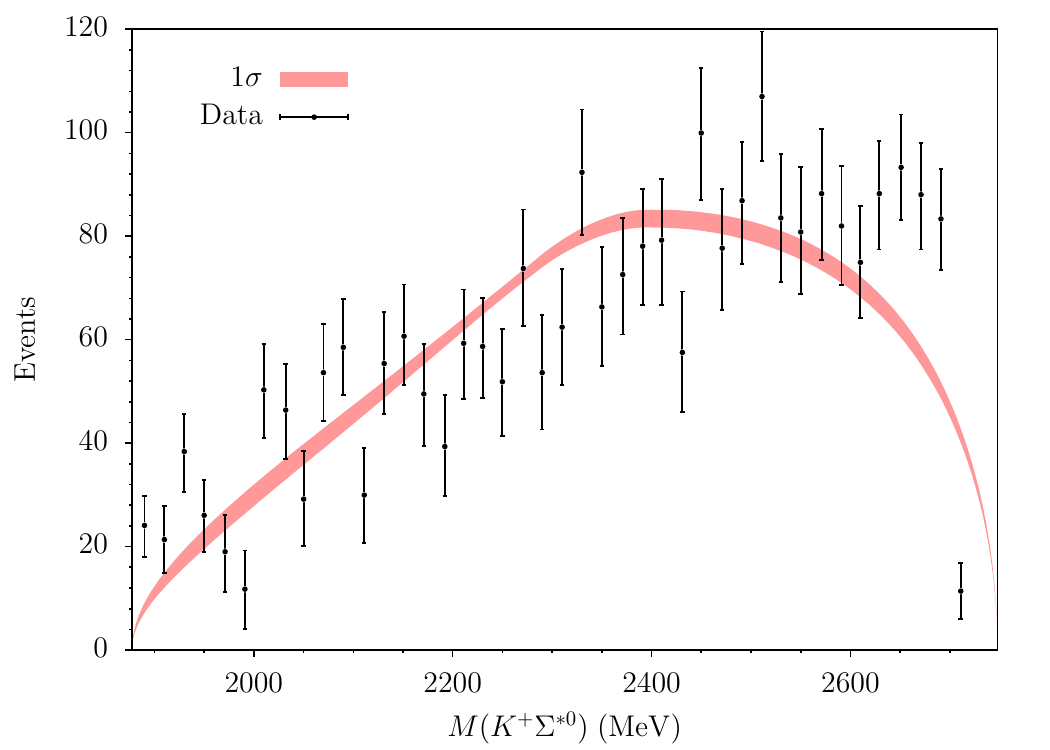}
		\label{fig:error_KSigmastar}
	}\hfill
	\subfigure[]{
		\includegraphics[scale=0.43]{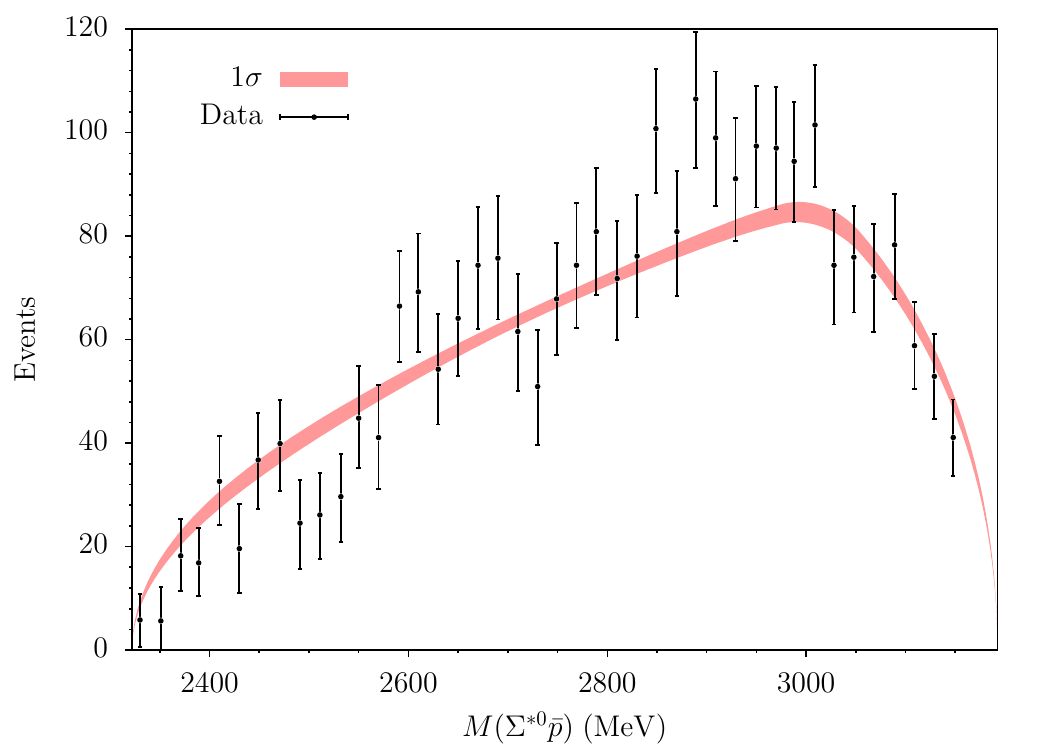}
		\label{fig:error_SigmastarPbar}
	}
	\caption{Theoretical uncertainties in the invariant mass distributions of $\bar{p}K^{+}$~(a), $K^{+}\Sigma^{*0}$~(b), $\Sigma^{*0}\bar{p}$~(c) for the $\psi(3686)\to\bar{p}K^+\Sigma^*(1385)$ process, estimated using the parametric bootstrap method. The ``Data" denotes the BESIII experimental data~\cite{BESIII:2026kuj}.}
	\label{fig:error_mass_distribution_BBP}
\end{figure}

For the $\psi(3686)\to\bar{p}K^{*0}(892)\Sigma^+$ decay, we fit the three free parameters $A$, $B$, and $B_t$ to the BESIII data~\cite{BESIII:2026kuj} within the kinematically allowed region. The fit yields $A=0.54$, $B=1.23$, and $B_t=-0.40$, with $\chi^2/\mathrm{d.o.f.}=398.24/91=4.38$.

Figure~\ref{fig:invariant_mass_distribution_BBV} shows the $\bar{p}K^{*0}$, $K^{*0}\Sigma^+$, and $\Sigma^+\bar{p}$ invariant mass distributions. The green dashed and blue dash-dotted curves represent the tree-level and vector--baryon final-state interaction (FSI) contributions, respectively. The black solid curves show the full distributions, including interference between the two amplitudes. The $\bar{p}K^{*0}$ spectrum exhibits an enhancement near the $\bar{K}^{*}N$ threshold at approximately $1830\,\mathrm{MeV}$. In the model, this structure arises from the coherent sum of the tree-level amplitude and the FSI amplitude associated with the dynamically generated $\Sigma(1830)$ state with $J^P=3/2^-$, predicted in Ref.~\cite{Oset:2010tof}.

The PDG lists three other resonances in this mass region: $\Sigma(1880)$ with $J^P=1/2^+$, $\Sigma(1900)$ with $J^P=1/2^-$, and $\Sigma(1910)$ with $J^P=3/2^-$. The $\Sigma(1880)$ has a width of approximately $200\,\mathrm{MeV}$, with no reported decay to $\bar{K}^{*}N$. The $\Sigma(1900)$ has a mass of approximately $1925\,\mathrm{MeV}$, above the observed enhancement, and a width of approximately $165\,\mathrm{MeV}$. Its reported branching fraction to $\bar{K}^{*}N$ is approximately $7\%$. The $\Sigma(1910)$ has a width of approximately $220\,\mathrm{MeV}$, and its only listed $\bar{K}^{*}N$ branching fraction is approximately $1\%$ for the $D$-wave contribution. The position of the enhancement and the reported decay properties of these nearby states favor the association of the experimental peak of the $K^*\bar{p}$ mass distribution with the dynamically generated $\Sigma(1830)$.

For the $\psi(3686)\to\bar{p}K^+\Sigma^{*0}(1385)$ decay, we fit the three free parameters $\tilde{A}$, $\tilde{B}$, and $\tilde{C}$ to the BESIII data~\cite{BESIII:2026kuj}, applying the same kinematic selection criterion. The fit yields $\tilde{A}=-1.06\times10^{-3}$, $\tilde{B}=1.64\times10^{-4}$, and $\tilde{C}=-5.20\times10^{-4}$, with $\chi^2/\mathrm{d.o.f.}=271.08/123=2.20$.

Figure~\ref{fig:invariant_mass_distribution_BBP} shows the $\bar{p}K^+$, $K^+\Sigma^{*0}$, and $\Sigma^{*0}\bar{p}$ invariant mass distributions. The red dashed curves represent the tree-level contribution, the blue dash-dotted curves show the $K^-p$ FSI contribution, and the black solid curves show the full distributions. The $\Sigma(1430)$, dynamically generated by the $K^-p$ interaction, produces an enhancement near the $\bar{p}K^+$ threshold. This enhancement is more pronounced in the isolated FSI contribution than in the full distribution.

We should mention that the amplitudes that we get with the chiral unitary approach~\cite{Oset:1997it} are reliable in a range of energies that do not exceed $1550$~MeV in the $\bar{p}K^+$  invariant mass. Thus, we do as it is usually done~\cite{Debastiani:2016ayp,Duan:2024czu,Dai:2026zqn}, killing adiabatically the amplitudes after a certain energy. We take
\begin{eqnarray}
	Gt(M_{\text{inv}})=Gt(M_{\text{cut}})e^{-\alpha(M_{\text{inv}}-M_{\text{cut}})} ~~~~\text{for}~M_{\text{inv}}>M_{\text{cut}},
\end{eqnarray}
with $M_{\text{cut}}=1550$~MeV, $\alpha=3.7\times10^{-3}$~MeV$^{-1}$ as done in \cite{Debastiani:2016ayp,Duan:2024czu,Dai:2026zqn}, where it was shown that moderate
changes of $M_{\text{cut}}$ and $\alpha$ have only minor effects on the mass distributions.

We use a parametric bootstrap to propagate the experimental uncertainties to the predicted invariant mass distributions~\cite{Efron:1986hys,Albaladejo:2016hae,Lyu:2026rsm,Lyu:2026mvu}. We generate $10^6$ pseudo-data sets by sampling each data point from a Gaussian distribution centered on its measured value, with a standard deviation equal to its quoted experimental uncertainty. For each pseudo-data set, we refit the three model parameters using the original experimental uncertainties and recalculate the invariant mass distributions. The shaded bands in Fig.~\ref{fig:error_mass_distribution_BBP} show the resulting pointwise $1\sigma$ statistical uncertainties. In the $\bar{p}K^+$ spectrum, the uncertainty band is slightly wider near threshold than at higher invariant masses. More precise measurements in the threshold region will help clarify the properties of the $\Sigma(1430)$. The uncertainty bands in Fig.~\ref{fig:invariant_mass_distribution_BBV} are similar.

\section{ Conclusions }\label{sec4}

We have investigated the decays $\psi(3686)\to\bar{p}K^{*0}(892)\Sigma^+$ and $\psi(3686)\to\bar{p}K^+\Sigma^{*0}(1385)$ in a coupled-channel framework. The decay amplitudes include both direct production and $S$-wave final state interaction $\bar{p}K^{*0}$ and $\bar{p}K^+$, respectively. The rescattering amplitudes are obtained from the corresponding isospin-one vector-meson--baryon and pseudoscalar-meson--baryon interactions. For each decay, we fit three production parameters to the BESIII data and compare the resulting distributions with the measured ones for all three final-state pairs.

In the $\bar{p}K^{*0}$ distribution, the coherent sum of direct production and rescattering associated with the dynamically generated $\Sigma(1830)$ produces an enhancement near $1830$~MeV, close to the nominal $\bar{K}^{*}N$ threshold. In the $\bar{p}K^+$ distribution, the rescattering contribution associated with the $\Sigma(1430)$ also exhibits a near-threshold enhancement, but this feature is less pronounced in the total distribution. The fits yield $\chi^2/\mathrm{d.o.f.}=4.38$ and $2.20$ for the vector- and pseudoscalar-meson decay modes, respectively. The model thus offers a common interpretation of the two enhancements in terms of meson--baryon rescattering, and the two dynamically generated resonances.

For the $\psi(3686)\to\bar{p}K^+\Sigma^{*0}(1385)$ decay, we used a parametric bootstrap to propagate the experimental uncertainties through the fit to the calculated invariant mass distributions. The resulting pointwise statistical uncertainty band in the $\bar{p}K^+$ distribution is slightly wider near threshold than at higher invariant masses. More precise measurements in both threshold regions would help constrain the relative contributions of direct production and rescattering and clarify the properties of the $\Sigma(1430)$ and $\Sigma(1830)$ resonances.

\section*{Acknowledgments}
This work was supported by the National Key R\&D Program of China (Grant No. 2024YFE0105200), the National Natural Science Foundation of China under Grant No. 12475086 and No. 12192263, the Natural Science Foundation of Henan (Grant No. 252300423951), and the Zhengzhou University Young Student Basic Research Projects for PhD students (Grant No. ZDBJ202522). Wen-Tao Lyu acknowledges the support of the China Scholarship Council. 
F.C.S. gratefully acknowledges the support received from the S$\tilde{\text{a}}$o Paulo Research Foun-
dation (FAPESP), Brazil, (Grant numbers: 2024/19103-9 and 2026/04916-0).
Man-Yu Duan is also supported by the Fundamental Research Funds for the Central Universities (No. 2026JKF02ZK14).
This work is also partly supported by the Spanish Ministerio de Economia y Competitividad~(MINECO) and European FEDER funds under Contracts No. FIS2017-84038-C2-1-PB and No. PID2020--112777 GB-I00 and by Generalitat Valenciana under Contract No. PROMETEO/2020/023. This project has received funding from the European Union Horizon 2020 research and innovation program under the program H2020-INFRAIA-2018-1, Grant Agreement No. 824093 of the STRONG-2020 project.
Research was partially supported by Grant No. PID2023-147458NB-C21 funded by MCIN/AEI/ 10.13039/501100011033 and by the European Union.

\end{document}